\documentclass{appolb}
\usepackage{epsfig}

\def\bge{\begin{equation}}
\def\ene{\end{equation}}
\def\bg{\begin{eqnarray}}
\def\en{\end{eqnarray}}

\def\ubar{{\bar{u}}}
\def\dbar{{\bar{d}}}
\def\sbar{{\bar{s}}}

\def\bge{\begin{equation}}
\def\ene{\end{equation}}
\def\bg{\begin{eqnarray}}
\def\en{\end{eqnarray}}

\def\ubar{{\bar{u}}}
\def\dbar{{\bar{d}}}
\def\sbar{{\bar{s}}}

\begin{document}
\title{QCD SYMMETRIES IN $\eta$ AND $\eta'$ MESIC NUCLEI
\thanks{Presented by SDB at the Second International Symposium on Mesic Nuclei, 
Cracow, September 22-25 2013.}
}
\author{Steven D. Bass 
\address{Stefan Meyer Institute for Subatomic Physics, \\
Austrian Academy of Sciences,
Boltzmanngasse 3, 1090 Vienna, Austria}
\and
Anthony W. Thomas
\address{CSSM and
ARC Centre of Excellence for Particle Physics at the Terascale, \\
School of Chemistry and Physics, 
University of Adelaide, \\
Adelaide SA 5005, Australia}
}
\maketitle
\begin{abstract}
We discuss the role of QCD symmetries and confinement in understanding $\eta$ and $\eta'$ mesic nuclei.
$\eta$ and $\eta'$ bound states in nuclei are sensitive to the flavour-singlet component in the meson. The bigger the singlet component, the more attraction and the greater the binding.
Recent results on 
the $\eta'$ mass in nuclei from the CBELSA/TAPS collaboration 
are very similar to the prediction of the Quark Meson Coupling model. 
In the model $\eta$-$\eta'$ mixing induces a factor of two enhancement of the $\eta$-nucleon scattering length $a_{\eta N}$ relative to 
the prediction with a pure octet $\eta$ with real part about 0.8 fm.
\end{abstract}

\section{Introduction}

Measurements of
the $\eta-$ and $\eta'-$ (as well as pion and kaon)
nucleus systems promise to yield valuable new information about
dynamical chiral and axial U(1) symmetry breaking in low energy QCD.
The quark condensate is modified in the nuclear environment
which leads to changes in 
the properties of hadrons in medium
including the masses of the Goldstone bosons 
\cite{kienle}.
While pions and kaons are would-be Goldstone bosons associated 
with chiral symmetry, the isosinglet $\eta$ and $\eta'$ mesons 
are too massive by about 300-400 MeV for them to be pure 
Goldstone states \cite{sbcracow}.
They receive extra mass from non-perturbative gluon dynamics associated with the QCD axial anomaly.
How does this gluonic part change in nuclei ?

The $\eta$ and $\eta'$-nucleon interactions are attractive suggesting that these mesons may form strong-interaction bound-states in nuclei.
Medium modifications can be understood at the quark level 
through coupling of the scalar isoscalar $\sigma$ 
(and also $\omega$ and $\rho$) mean fields in the nucleus 
to the light quarks in the hadron \cite{qmc}.
The binding energies and in-medium masses of the $\eta$ and $\eta'$
are sensitive to the flavour-singlet component in the mesons and 
hence to the non-perturbative glue associated with axial U(1) dynamics \cite{etaA}.
There is presently a vigorous experimental programme to search 
for evidence of these bound states with ongoing experiments at 
COSY to look for possible $\eta$ bound states in Helium 
\cite{pawela}
and new experiments at ELSA \cite{elsap}
and GSI/FAIR \cite{gsi}
to look for possible $\eta'$ bound states in Carbon.

In this paper we discuss the relation between QCD symmetries and 
possible medium effects in these mesons.
Important considerations are confinement, 
chiral and axial U(1) symmetry, 
their behaviour in medium and the role of $\eta$-$\eta'$ mixing.
We focus on the 
Quark Meson Coupling model (QMC, for a review see \cite{qmc}) 
and compare its predictions 
for the $\eta$ and $\eta'$ systems
to chiral coupled channels, NJL and linear $\sigma$ models.

Meson masses in nuclei are determined from the meson nucleus optical potential and the scalar induced contribution 
to the meson propagator evaluated at zero three-momentum, 
${\vec k} =0$, in the nuclear medium.
Let $k=(E,{\vec k})$ and $m$ denote the four-momentum and mass of the meson in free space.
Then, one solves the equation
\begin{equation}
k^2 - m^2 = {\tt Re} \ \Pi (E, {\vec k}, \rho)
\end{equation}
for ${\vec k}=0$
where $\Pi$ is the in-medium $s$-wave meson self-energy.
Contributions to the in medium mass come from coupling to the scalar 
$\sigma$ field in the nucleus in mean-field approximation,
nucleon-hole and resonance-hole excitations in the medium.
For ${\vec k}=0$, $k^2 - m^2 \sim 2 m (m^* - m)$ 
where $m^*$ is the effective mass in the medium.
The mass shift $m^*-m$
is the depth or real part of the meson nucleus optical potential.
The imaginary part of the potential measures the width of the meson
in the nuclear medium.
The $s$-wave self-energy can be written as \cite{ericson}
\begin{equation}
\Pi (E, {\vec k}, \rho) \bigg|_{\{{\vec k}=0\}}
=
- 4 \pi \rho \biggl( { b \over 1 + b \langle {1 \over r} \rangle } \biggr) .
\end{equation}
Here $\rho$ is the nuclear density,
$
b = a ( 1 + {m \over M} )
$
where 
$a$ is the meson-nucleon scattering length, 
$M$ is the nucleon mass and
$\langle {1 \over r} \rangle$ is the inverse correlation length,
$\langle {1 \over r} \rangle \simeq m_{\pi}$ 
for nuclear matter density.
Attraction corresponds to positive values of $a$.
The denominator in Eq.(2) is the Ericson-Ericson-Lorentz-Lorenz
double scattering correction.

What should we expect for the $\eta$ and $\eta'$ ?

COSY searches are focussed on possible $\eta$ bound states in 
$^3$He and $^4$He \cite{pawela}.
Eta bound states in Helium require a large $\eta-$nucleon 
scattering length 
with real part greater than about 0.9 fm \cite{gal}.
For clean observation of a bound state one needs the real part 
of the optical potential to be much bigger than the imaginary part.
We refer to \cite{niskanen} for 
a recent discussion of the $\eta$ in light nuclei.

For the $\eta'$ new experiments are planned to look for possible bound states in Carbon
using the (p, d) reaction at GSI \cite{gsi} 
and in photoproduction at ELSA \cite{elsap}.
Recent measurements of the transparency ratio for $\eta'$ photoproduction from nuclear targets have been interpreted 
to mean a small $\eta'$ width in nuclei
$20 \pm 5.0$ MeV at nuclear matter density $\rho_0$ \cite{elsa}
that might give rise to relatively narrow bound 
$\eta'$-nucleus states accessible to experiments.

In addition to bound state searches, meson mass shifts 
can also be investigated through studies of excitation 
functions in photoproduction experiments from nuclear targets. 
The production cross section is enhanced with the lower effective mass in the nuclear medium. 
When the meson leaves the nucleus it returns on-shell 
to its free mass with the energy budget conserved at
the expense of the kinetic energy so that excitation functions
and momentum distributions can provide essential clues to the
meson properties in medium \cite{metag}.
Using this physics a first (indirect) estimate of the $\eta'$ 
mass shift has recently been deduced 
by the CBELSA/TAPS Collaboration \cite{nanova}.
The $\eta'$-nucleus optical potential 
$V_{\rm opt} = V_{\rm real} + iW$
deduced from these photoproduction experiments is 
\begin{eqnarray}
V_{\rm real} (\rho_0)
= m^* - m 
&=& -37 \pm 10 (stat.) \pm 10 (syst.) \ {\rm MeV}
\nonumber \\ 
W(\rho_0) &=& -10 \pm 2.5 \ {\rm MeV}
\end{eqnarray}
at nuclear matter density.
These numbers suggest that
bound states may be within reach of forthcoming experiments.
The mass shift, Eq.(3), is also very similar 
to the expectations of the Quark Meson Coupling model, see below.

\section{QCD symmetries and the $\eta$ and $\eta'$}

Low energy QCD is characterised by confinement and dynamical
chiral symmetry breaking.
The absence of parity doublets in the hadron spectrum tells 
us that the near-chiral symmetry for light $u$ and $d$ quarks 
is spontaneously broken. 
Scalar confinement implies dynamical chiral symmetry breaking. 
For example, in the Bag model the Bag wall connects left and right
handed quarks 
leading to quark-pion coupling and the pion cloud of the nucleon
\cite{awtcbm}.
Spontaneous chiral symmetry breaking in QCD is associated with a 
non-vanishing chiral condensate
\footnote{
Spontaneous symmetry breaking and phase transitions give changes 
in the vacuum energy which couples to gravity. 
It is interesting that the QCD quark condensate gives a 
contribution $10^{44}$ times larger
(and the electroweak Higgs condensate $10^{56}$ times larger)
than the nett vacuum energy extracted from the cosmological 
constant without additional physics, see {\it e.g.} \cite{SBober}.
Further, the vacuum energy extracted from the cosmological constant
is small and positive whereas the QCD and Higgs contributions come with a negative sign.
}
\begin{equation}
\langle \ {\rm vac} \ | \ {\bar \psi} \psi \ | \ {\rm vac} \ \rangle < 0
.
\label{eq5}
\end{equation}
This spontaneous symmetry breaking induces an octet of 
Goldstone bosons associated with SU(3) and also 
(before extra gluonic effects in the singlet channel)
a flavour-singlet Goldstone boson.
The Goldstone bosons $P$ couple to the axial-vector currents 
which play the role of Noether currents through
\begin{equation}
\langle {\rm vac} | J_{\mu 5}^i | P(p) \rangle = 
-i f_P^i \ p_{\mu} e^{-ip.x}
\end{equation}
with $f_P^i$ the corresponding decay constants
and satisfy the Gell-Mann-Oakes-Renner relation
\begin{equation}
m_{\pi}^2 f_{\pi}^2 = - m_q \langle {\bar \psi} \psi \rangle .
\end{equation}
The pion and kaon fit well in this picture.

The isosinglet $\eta$ and $\eta'$ masses are about 300-400 MeV 
too heavy to be pure Goldstone states.
One needs extra mass in the flavour-singlet channel associated 
with non-perturbative topological gluon configurations 
\cite{Shore,crewther}, 
related perhaps to confinement \cite{ks} or instantons \cite{thooft}.
SU(3) breaking generates mixing between the octet and 
singlet states yielding the massive $\eta$ and $\eta'$ bosons.

With increasing density chiral symmetry is partially restored corresponding to a reduction in the value of the quark 
condensate and pion decay constant $f_{\pi}$.
Experiments with pionic atoms give a value
$f_{\pi}^{*2}/f_{\pi}^2 = 0.64 \pm 0.06$ 
at nuclear matter density $\rho_0$ \cite{kienle}.
This suggests changes in the meson masses in the medium 
and 
also the coupling of the Goldstone bosons to the constituent 
quarks and the nucleon
-- 
for reviews of 
the QCD phase diagram 
and medium modifications at 
finite density see \cite{wambach,metagrev,qmc}.
These medium modifications need to be understood 
self-consistently within the interplay of confinement,
spontaneous chiral symmetry breaking and axial U(1)
dynamics.
In the limit of chiral restoration the pion should decouple 
from the physics, the pion decay constant 
$f_{\pi}$ go to zero and
(perhaps) with scalar confinement 
the pion constituent-quark and pion nucleon coupling
constants should vanish with dissolution of the pion wavefunction.

To see the effect of the gluonic mass contribution consider the 
$\eta$-$\eta'$ mass matrix for free mesons
(at leading order in the chiral expansion) 
\begin{equation}
M^2 =
\left(\begin{array}{cc}
{4 \over 3} m_{\rm K}^2 - {1 \over 3} m_{\pi}^2  &
- {2 \over 3} \sqrt{2} (m_{\rm K}^2 - m_{\pi}^2) \\
\\
- {2 \over 3} \sqrt{2} (m_{\rm K}^2 - m_{\pi}^2) &
[ {2 \over 3} m_{\rm K}^2 + {1 \over 3} m_{\pi}^2 + {\tilde m}^2_{\eta_0} ]
\end{array}\right)
.
\label{eq10}
\end{equation}
Here ${\tilde m}^2_{\eta_0}$ is the flavour-singlet gluonic mass term.

The masses of the physical $\eta$ and $\eta'$ mesons are found
by diagonalizing this matrix, {\it viz.}
\begin{eqnarray}
| \eta \rangle &=&
\cos \theta \ | \eta_8 \rangle - \sin \theta \ | \eta_0 \rangle
\\ \nonumber
| \eta' \rangle &=&
\sin \theta \ | \eta_8 \rangle + \cos \theta \ | \eta_0 \rangle
\label{eq11}
\end{eqnarray}
where
\begin{equation}
\eta_0 = \frac{1}{\sqrt{3}}\; (u\ubar + d\dbar + s\sbar),\quad
\eta_8 = \frac{1}{\sqrt{6}}\; (u\ubar + d\dbar - 2 s\sbar) 
.
\label{mixing2}
\end{equation}
One obtains values for the $\eta$ and $\eta'$ masses:
\begin{equation}
m^2_{\eta', \eta} 
= (m_{\rm K}^2 + {\tilde m}_{\eta_0}^2 /2)
\pm {1 \over 2}
\sqrt{(2 m_{\rm K}^2 - 2 m_{\pi}^2 - {1 \over 3} 
{\tilde m}_{\eta_0}^2)^2 + {8 \over 9} {\tilde m}_{\eta_0}^4} 
.
\label{eq12}
\end{equation}

The gluonic mass term is obtained from summing over the two 
eigenvalues in Eq.(10)
\begin{equation}
m_{\eta}^2 + m_{\eta'}^2 = 2 m_K^2 + {\tilde m}_{\eta_0}^2 .
\end{equation}
Substituting the physical values of 
$m_{\eta}$, $m_{\eta'}$ and  $m_K$ 
gives ${\tilde m}_{\eta_0}^2 = 0.73$GeV$^2$ \cite{vecca}.
The gluonic mass term has a rigorous interpretation in terms of
the Yang Mills topological susceptibility, 
see {\it e.g.} \cite{Shore,venez}.

In the OZI limit of no gluonic mass term
the $\eta$ would be approximately an isosinglet light-quark state
(${1 \over \sqrt{2}} | {\bar u} u + {\bar d} d \rangle$)
with mass $m_{\eta} \sim m_{\pi}$
degenerate with the pion and
the $\eta'$ would be a strange-quark state $| {\bar s} s \rangle$
with mass $m_{\eta'} \sim \sqrt{2 m_{\rm K}^2 - m_{\pi}^2}$
--- mirroring the isoscalar vector $\omega$ and $\phi$ mesons.

Phenomenological studies of various decay processes give a value 
for the $\eta$-$\eta'$ mixing angle between 
$-15^\circ$ and $-20^\circ$ \cite{gilman}.
This mixing means that non-perturbative glue through axial U(1)
dynamics plays an important role in both the $\eta$ and $\eta'$ 
and their interactions \cite{sbcracow}.
Treating the $\eta$ as an octet 
pure would-be Goldstone boson risks losing essential physics.
Processes associated with the flavour-singlet $1^{-+}$ channel 
are characterized by OZI violation.
The anomalous glue that generates the large $\eta$ and $\eta'$ 
masses also drives OZI violating $\eta$ and $\eta'$ production 
and decay processes [24-28],
enters in the $\eta'$-nucleon interaction \cite{bass99}
and the flavour-singlet Goldberger-Treiman relation 
\cite{tgv}
associated with the proton spin puzzle \cite{SBrmp1,SBrmp2}.

The gluonic mass term is related to the QCD axial anomaly 
in the divergence of the flavour-singlet axial-vector current.
While the non-singlet axial-vector currents are partially conserved (they have just mass terms in the divergence), the singlet current
$
J_{\mu 5} = \bar{u}\gamma_\mu\gamma_5u
+ \bar{d}\gamma_\mu\gamma_5d + \bar{s}\gamma_\mu\gamma_5s 
$
satisfies the anomalous divergence equation 
\begin{equation}
\partial^\mu J_{\mu5} = 6 Q
+ \sum_{k=1}^{3} 2 i m_k \bar{q}_k \gamma_5 q_k 
\end{equation}
where 
$
Q = \partial^{\mu} K_{\mu}
= {\alpha_s \over 8 \pi} G_{\mu \nu} {\tilde G}^{\mu \nu}
$
is the topological charge density.
The integral over space $\int \ d^4 z \ Q = n$ measures the 
gluonic winding number \cite{crewther} 
which is an integer for (anti-)instantons and which vanishes 
in perturbative QCD.

Within the low energy effective chiral Lagrangian for QCD
the gluonic mass term is introduced via a flavour-singlet
potential involving the topological charge density $Q$
which is constructed so that the Lagrangian also reproduces 
the axial anomaly \cite{vecca}.
In this approach the medium dependence of 
${\tilde m}_{\eta_0}^2$ 
is introduced through coupling to the $\sigma$ 
(correlated two-pion) mean-field in the nucleus
through the interaction term
$
{\cal L}_{\sigma Q} = g_{\sigma Q} \ Q^2 \ \sigma
$
where 
$g_{\sigma Q}$ denotes coupling to the $\sigma$ mean field.
One finds the gluonic mass term decreases in-medium
${\tilde m}_{\eta_0}^{*2} < {\tilde m}_{\eta_0}^2$ 
independent of the sign of $g_{\sigma Q}$ and the medium acts
to partially neutralise axial U(1) symmetry breaking by gluonic effects \cite{etaA}.

To estimate the size of the effect we look to phenomenology and
QCD motivated models.
QCD inspired models of the $\eta$ and $\eta'$ nucleus systems are
constructed with different selections of 
``good physics input'':
how they treat confinement, chiral symmetry and axial U(1) dynamics.

\section{The $\eta$ and $\eta'$ in nuclei}

The physics of $\eta$ and $\eta'$ mass shifts with $\eta$-$\eta'$ mixing has been investigated 
by Bass and Thomas \cite{etaA,etaApol}
within the Quark Meson Coupling model 
of hadron properties in the nuclear medium \cite{qmc,etaqmc}.
Here the large $\eta$ and $\eta'$ masses are used to motivate 
taking an MIT Bag description for the meson wavefunctions.
Gluonic topological effects are understood to be ``frozen in'',
meaning that they are only present implicity through the masses
and mixing angle in the model.
The in-medium mass modification comes from coupling the light 
(up and down) quarks and antiquarks in the meson wavefunction 
to the scalar $\sigma$ mean-field in the nucleus 
working in mean-field approximation \cite{qmc}.
The coupling constants in the model for the coupling of light-quarks
to the $\sigma$ (and $\omega$ and $\rho$) mean-fields in the nucleus
are adjusted to fit the saturation energy and density of
symmetric nuclear matter and the bulk symmetry energy.
The strange-quark component of the wavefunction does not couple
to the $\sigma$ field and $\eta$-$\eta'$ mixing is readily built 
into the model.
Gluon fluctuation and centre-of-mass effects are assumed to be independent of density.
The model results for the meson masses in medium and the real 
part of the meson-nucleon scattering lengths
are shown in Table 1 for different values of 
the $\eta$-$\eta'$ mixing angle, 
which is taken to be density independent.
\footnote{
The values of ${\tt Re} a_{\eta}$ quoted in Table 1 are obtained
from substituting the in-medium and free masses into Eq.(2) with
the Ericson-Ericson denominator turned-off
(since we choose to work in mean-field approximation).}
The QMC model makes no claim about the imaginary part of the scattering length.

Increasing the flavour-singlet component in the $\eta$ at 
the expense of the octet component gives more attraction, 
more binding and a larger value of the $\eta$-nucleon scattering length, $a_{\eta N}$. 
Since the mass shift is approximately proportional to the 
$\eta$--nucleon scattering length, it follows that that 
the physical value of $a_{\eta N}$ should be larger than 
if the $\eta$ were a pure octet state.
For the $\eta'$ the opposite is true: the greater the mixing angle
the smaller the singlet component in the $\eta'$ and smaller 
the value of the $\eta'$ binding energy and the 
$\eta'$-nucleon scattering length.

\begin{table}[t!]
\begin{center}
\caption{
Physical masses fitted in free space, the bag masses in medium at normal nuclear-matter density, $\rho_0 = 0.15$ fm$^{-3}$, 
and corresponding meson-nucleon scattering lengths.
}
\label{bagparam}
\begin{tabular}[t]{c|lll}
\hline
&$m$ (MeV) 
& $m^*$ (MeV) & ${\tt Re} a$ (fm)
\\
\hline
$\eta_8$  &547.75  
& 500.0 &  0.43 \\
$\eta$ (-10$^o$)& 547.75  
& 474.7 & 0.64 \\
$\eta$ (-20$^o$)& 547.75  
& 449.3 & 0.85 \\
$\eta_0$  &      958 
& 878.6  & 0.99 \\
$\eta'$ (-10$^o$)&958 
& 899.2 & 0.74 \\
$\eta'$ (-20$^o$)&958 
& 921.3 & 0.47 \\
\hline
\end{tabular}
\end{center}
\end{table}

There are several key observations.

\subsection{$\eta$ bound states}

$\eta$-$\eta'$ mixing with the phenomenological mixing angle 
$-20^\circ$ leads to a factor of two increase 
in the mass-shift and 
in the scattering length obtained in the model
relative to the prediction for a pure octet $\eta_8$.
This result may explain why values of $a_{\eta N}$ extracted from 
phenomenological fits to experimental data where the $\eta$-$\eta'$ 
mixing angle is unconstrained 
give larger values than those predicted 
in theoretical models where the $\eta$ is treated as a pure octet state.

COSY-11 have studied the final state interaction (FSI)
in measurements of $\eta$ production in proton-proton 
collisions close to threshold
based on the effective range approximation.
The results are consistent with the scattering length
$a_{\eta N} \simeq 0.7$ + i 0.4fm \cite{cosyeta}.
K-matrix fits to experimental data suggest a value close 
to 0.9 fm for the real part of $a_{\eta N}$ \cite{wycech}.
In contrast, smaller values of $a_{\eta N}$ with real part 
$\sim 0.2$ fm are predicted by chiral coupled channels where 
the $\eta$ meson is treated in pure octet approximation 
\cite{etaweise,etaoset}. 
These chiral models involve performing a coupled channels analysis of $\eta$ and $\eta'$ production after multiple rescattering in the nucleus with potentials obtained from the SU(3) chiral Lagrangian.

For the kaon nucleus system similar results are obtained using 
the QMC and chiral coupled channels models,
e.g. reduction in the $K^-$ mass of about 100 MeV at nuclear matter
density and interpretation of the $\Lambda (1405)$ resonance
as dynamically generated in the kaon-nucleon system
\cite{qmckaon,etaweise}.
An effective mass drop of the $K^-$ to about 270 MeV at 
$2 \rho_0$ 
has been observed in heavy-ion collisions \cite{kaons}, 
consistent with the predictions of the Quark Meson Coupling
and chiral coupled channels model calculations.

The $\eta$-nucleon interaction is characterised by a strong coupling 
to the $S_{11}$(1535) nucleon resonance.
For example, $\eta$ meson production in proton nucleon collisions 
close to threshold is known to procede via a strong isovector exchange contribution with excitation of the $S_{11}(1535)$.
Recent measurements of $\eta'$ production suggest a different
mechanism for this meson \cite{pawel}. 
Experiments in heavy-ion collisions \cite{averbeck} and $\eta$ photoproduction from nuclei \cite{robig,yorita} suggest little modification of 
the $S_{11} (1535)$ excitation in-medium.
In quark models the $S_{11}$ is interpreted as a 3-quark state: 
$(1s)^2(1p)$.
In QMC the excitation energy is $\sim 1544$ MeV, consistent
with observations,
with the scalar attraction compensated by repulsion 
from coupling to the $\omega$ mean-field to give the excitation energy \cite{etaA}.
Small mass shift is also reported in the coupled channels models 
where the $S_{11}$ is instead interpreted as a $K \Sigma$ 
quasi-bound state, with the $\eta$ instead treated as a pure octet state \cite{kaiser}.

\subsection{$\eta'$ bound states}

From Table 1 the QMC prediction for the $\eta'$ mass in medium at
nuclear matter density is 921 MeV with mixing angle of $-20^\circ$.
This value is in excellent agreement with the mass shift 
$-37 \pm 10 \pm 10$ MeV
deduced from photoproduction data \cite{nanova}.
For the $\eta'$-nucleon scattering length a conservative 
upper bound 
$| {\tt Re} a_{\eta' N} | < 0.8$fm 
has been deduced by COSY-11 \cite{cosy}
by comparing the FSI in $\pi^0$ and $\eta'$ 
production in proton-proton collisions close to threshold.

If we assume the mass formula (11) holds also in symmetric 
nuclear matter at finite density and substitute
the QMC predictions for the $\eta$, $\eta'$ and 
kaon masses in medium ($m_K^* = 430.4$ MeV), 
then we obtain 
$
{\tilde m}_{\eta_0}^{*2}(\rho_0) 
= 0.68 {\rm GeV}^2 < 0.73 {\rm GeV}^2
$
with $\eta$-$\eta'$ mixing angle equal to $-20^\circ$.

Mixing increases the octet relative to singlet component in
the $\eta'$, reducing the binding through increased strange
quark component in the $\eta'$ wavefunction.
Without the gluonic mass contribution, the $\eta'$ 
would be a strange quark state after $\eta$-$\eta'$ mixing.
Within the QMC model there would be no coupling 
to the $\sigma$ mean field and 
no mass shift, 
so that any observed mass shift is induced 
by the QCD gluon anomaly that generates part of the $\eta'$ mass.

New coupled channels model calculations have appeared 
with mixing and 
vector meson channels included, fitted to a range of
possible values of $a_{\eta' N}$ \cite{osetetaprime}.
In both QMC and coupled channels calculations the $\eta$ and 
$\eta'$ masses decrease in medium and confinement is inbuilt 
in the models, either through the Bag wavefunction with QMC
or using hadron degrees of freedom in coupled channels models.

In an alternative approach, large binding energies for the $\eta'$ 
up to 150 MeV were obtained 
in recent chiral NJL model calculations \cite{hirenzakinjl}. 
In this model the medium dependence of the $\eta$ mass 
depends strongly on the axial U(1) breaking.
The $\eta$ mass (and also pion mass) increases 
in medium as chiral symmetry is partially restored 
without axial U(1) breaking
or with density independent axial U(1) breaking.
Any decreasing $\eta$ mass depends on the density dependence of 
the axial U(1) breaking term which is put in through instantons. 
There is no confinement in the model and nuclear matter is treated 
as a Fermi gas of quarks (instead of a Fermi gas of nucleons).
Similar results for the mass shifts 
(about 100 MeV reduction for the $\eta'$ 
 and 50 MeV increase for the $\eta$)
were also found in linear sigma model based calculations 
in an hadronic basis
with the difference between the $\eta'$ and $\eta$ masses 
taken to be proportional to the quark condensate \cite{jido,jido2}.

These different theoretical results raise interesting questions
about the role of confinement, 
the representation of chiral symmetry (whether the $\sigma$
is better treated as a two pion resonance or as a quark-antiquark state) and 
how massive the light $0^-$ states 
can be 
for their wavefunctions to 
be treated as pure Goldstone bosons in the models.

\section{Summary and Open Questions}

Medium modifications of hadron properties are determined by chiral and flavour symmetries in QCD.
The $\eta$ and $\eta'$ are sensitive to flavour-singlet axial U(1) degrees of freedom.
QCD inspired models including confinement, chiral and axial U(1)
dynamics yield a range of predictions for the $\eta$ and $\eta'$
binding in nuclei.
The QMC prediction for 
the $\eta'$ mass shift is very similar to the recent value 
determined by CBELSA/TAPS from photoproduction experiments,
and the real part of the $\eta$-nucleon scattering length 
to values extracted from phenomenological fits to low-energy scattering data.
The positive results for the $\eta$ and $\eta'$ 
suggest one might also look at charmed mesons in nuclei,  
e.g. charmed mesic nuclei \cite{charmqmc}
where large binding energies are also predicted.
New data on possible $\eta$ and $\eta'$ bound states is 
expected soon from running and planned experiments at 
COSY, ELSA and GSI, and will help to pin down 
the dynamics of axial U(1) symmetry breaking in low-energy QCD.

\vspace{0.5cm}

{\bf Acknowledgements} \\

We thank A. Gal, 
S. Hirenzaki, D.  Jido, V. Metag, P. Moskal, M. Nanova,
K. Suzuki, M. Takizawa and K. Tsushima for helpful communications.
S.D.B. thanks
P. Moskal for the invitation to talk at this stimulating meeting. 
The research of S.D.B. is supported by the Austrian Science Fund, 
FWF, through grant P23753, and the Austrian Academy of Sciences,
while A.W.T. is supported by the Australian Research Council through
an Australian Laureate Fellowship (FL0992247) and 
the ARC Centre of Excellence in Particle Physics at the Terascale
and by the University of Adelaide .

\end{document}